\documentclass[superscriptaddress,twocolumn,prl,aps,showpacs]{revtex4-1}
\usepackage{epsfig}
\usepackage{bm,hyperref}
\usepackage{amssymb}
\usepackage{mathrsfs}
\usepackage{amsmath}
\usepackage{CJK}

\begin{document}

\title{Visualization of Dimensional Effects in Collective Excitations of Optically Trapped Quasi-Two-Dimensional Bose Gases}
\author{Ying Hu}
\email{yinggrant@gmail.com}
\affiliation{International Center for Quantum Materials, Peking University,
Beijing 100871, China}
\author{Zhaoxin Liang}
\email{zhxliang@gmail.com}
\affiliation{Shenyang National Laboratory for Materials Science, Institute of Metal
Research, Chinese Academy of Sciences, Wenhua Road 72, Shenyang 110016, China}

\date{\today}
\begin{abstract}
We analyze the macroscopic dynamics of a Bose gas axially confined in an optical lattice with a superimposed harmonic trap, taking into account weak tunneling effect. Our results show that upon transition to the quasi-two-dimensional (2D) regime of the trapped gas, the 3D equation of state and equilibrium density profile acquire corrections from 2D many-body effects. The corresponding frequency shift in the transverse breathing mode is accessible within current facilities, suggesting a direct observation of dimensional effects. Comparisons with other relevant effects are also presented.
\end{abstract}


\maketitle

The interplay between dimensionality and quantum fluctuations in two-dimensional (2D) strongly correlated quantum systems \cite{OLrev} has long been recognized to give rise to remarkable phenomena like high-$T_c$ superconductivity \cite{TopQC} and the long-sought Berezinskii-Kosterlitz-Thouless (BKT) transition \cite{BKT}. Recent extraordinary realization of quasi-2D ultracold Bose gases \cite{Q2DExp}, where tightly confined axial kinematics manifests as 2D features in pair collisions \cite{Petrov}, has cast new light in understanding low-dimensional many-body systems \cite{2D}.

In view of the important role played by dimensionality, their spectroscopic diagnostics is highly interesting \cite{DalfovoRev}. The interest particularly originates from the sensitivity of collective frequencies to equation of state which henceforth establishes their measurements as precise tests to the many-body physics. For example, the universal breathing mode of a 2D harmonically trapped Bose gas with $g\delta^2({\bf r})$ interaction, first noticed via the Castin-Dum-Kagan-Surkov-Shlyapnikov (CDKSS) scaling ansatz \cite{Scaling1,Scaling2}, reveals the hidden Pitaevskii-Rosch symmetry (PRS) \cite{Hsym,Chevy} in the associated classical field theory. Whereas, the frequency shift in this mode can provide a signature of quantum anomaly emerging upon quantization when quantum fluctuations significantly modify the scattering length \cite{Olshanii}.

\begin{figure}[tbh]
\includegraphics[width=6.0cm]{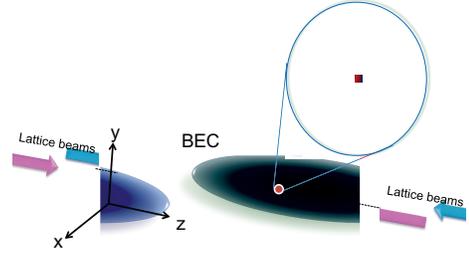}\label{Fig1}
\caption{(color online). Schematic picture of an optically trapped quasi-2D Bose gas. A 1D optical lattice is along the horizontal axis (z axis) with a BEC in an elongated harmonic trap, with axial (radial) frequency
$\omega_z$ ($\omega_{\perp}$). The BEC is thus confined to an array of narrow potential pancakes.}
\label{fig1}
\end{figure}

In this Letter, we are inspired to discuss visualization of the effect of dimensionality in collective excitations of an experimentally favored quasi-2D optically trapped Bose gas (seen in Fig. 1), taking the quantum tunneling into account. Collective oscillations in the presence of optical lattice has been previously studied intensively for the effective Hamiltonian $H_0=\sum_j\left[\frac{p^2_{j\perp}}{2m}+\frac{p^2_{jz}}{2m^{*}}\right]+\frac{m}{2}\sum_j\left[\omega^2_{\perp}\left(x_j^2+y_j^2\right)+\omega_z^2z_j^2\right]+\tilde{g}\sum_{j<k}\delta^3\left({\bf r}_{jk}\right)$ where $\tilde{g}$ is the lattice-renormalized 3D coupling constant and $m^*$ is the effective mass \cite{Kramer,Orso}. The corresponding superfluid hydrodynamic analysis \cite{Kramer} predicted that axial optical potential mark no effect on transverse modes for an elongated trap, which was later experimentally tested by Fort {\it et al} \cite{Fort}. However, this scenario will experience fundamental modifications when transiting to the quasi-2D regime. As we shall show in this Letter, upon the onset of ``frozen" axial kinematics in tight optical lattice, a 2D-peculiar position-dependent correction $H_1=\sum_{j<k}\tilde{g}_{1}\left({\bf r}_j\right)\delta^3\left({\bf r}_{jk}\right)$ emerges from the perspective of many-body theories \cite{Petrov}. Its impression on collective frequencies can be foreseen, for example, from the equation of motion
\begin{eqnarray}\label{Breath}
&&\frac{d^2F_B}{dt^2}+\left(2\omega_{\perp}\right)^2F_B=\frac{4}{m}\Big[H_0-\sum_j\Big(\frac{p^2_{jz}}{2m^*}+\frac{m}{2}\omega_z^2z_j^2\Big)\nonumber\\
&&-\sum_{j<k}\Big\{{\bf r}_j\cdot\nabla\tilde{g}_{1}\left({\bf r}_j\right)+\left[\nabla\tilde{g}_{1}\left({\bf r}_j\right)\right]\cdot{\bf r}_j\Big\}\delta^3\left({\bf r}_{jk}\right)\Big]
\end{eqnarray}
for the excitation operator $F_{B}=\sum_i (x_i^2+y_i^2)$ of the transverse breathing mode. Equation (\ref{Breath}), being equally valid classically as it is quantum mechanically, immediately leads to following statements: (i) for a pure 2D system within the classical field description, only $H_0$ survives on the right side of Eq. (\ref{Breath}) and the mode operator oscillates universally with $2\omega_{\perp}$, as required by PRS \cite{Hsym}; (ii) this universal oscillation with $2\omega_\perp$ also persists in a very elongated 3D dilute Bose gas described by $H_0$ ($\omega_z/\omega_\perp \ll 1$), as predicted by Ref. \cite{Kramer}; (iii) the oscillation frequency can be shifted from $2\omega_\perp$ by the emerging correction $\tilde{g}_1(\bf r)$.

In what follows, we analytically calculate the collective excitations of a quasi-2D Bose gas tightly confined by an optical lattice $V_{opt}=sE_R\sin^2(q_Bz)$, with a superimposed cylindrically symmetric harmonic trap $V_{ho}\left({\bf r}\right)=\frac{m}{2}(\omega^2_{\perp}x^2+\omega^2_{\perp}y^2+\omega_z^2z^2)$, as shown in Fig. 1. The lattice period is fixed by $q_B=\pi/d$ with $d$ being the lattice spacing, s is a dimensionless factor labeled by the intensity of a laser beam and $E_R=\hbar^2q^2_B/2m$ is the recoil energy with $\hbar q_B$ being the Bragg momentum.

Our starting point is the linearized hydrodynamic equation for density fluctuations $\delta n({\bf r},t)$, generalized straightforwardly from Ref.  \cite{Kramer} to the quasi-2D regime,
\begin{equation}\label{HD33}
m\frac{\partial ^2 \delta n}{\partial t^2}-\tilde{\nabla}\cdot\left[n\tilde{\nabla}\left(\frac{\partial \mu_{Q2D}}{\partial n}\delta n\right)\right]=0,
\end{equation}
with $\tilde{\nabla}\equiv\left(\nabla_{\perp},\nabla_z\sqrt{m/m^{*}}\right)$. The key ingredient of Eq. ({\ref{HD33}}) is the zero-temperature local chemical potential $\mu_{Q2D}$ of the quasi-2D Bose gas under consideration. Equation ({\ref{HD33}}) is justified by sufficiently weak tunneling which is nevertheless nonnegligible to ensure full coherence of the order parameter between different wells \cite{Kramer}, and by assuming the Thomas-Fermi (TF) limit and local density approximation \cite{DalfovoRev}. Here, the 3D density  $n\left({\bf r}\right)$ is determined from $\mu_0=\mu_{Q2D}\left[n({\bf r})\right]+V_{ho}\left({\bf r}\right)$, where $\mu_0$ is the ground state value of the chemical potential, fixed by the proper normalization of $n\left({\bf r}\right)$.

At the core of hydrodynamic analysis on collective oscillations is the knowledge of the equation of state. In order to determine $\mu_{Q2D}$ in Eq. ({\ref{HD33}}), we start from the grand partition function \cite{Popov} $Z=\int D\left[\psi^{*},\psi \right] e^{-{S\left[\psi^{*},\psi\right]}/{\hbar}}$ of an optically trapped quasi-2D Bose gas in the absence of harmonic potential, where
\begin{eqnarray}
S\left[\psi^{*},\psi\right]&=&\int d\tau \int
d^3\mathbf{r}\psi^{*} (\mathbf{r},\tau )\Bigg[\hbar \frac{\partial
}{\partial \tau }-
\frac{\hbar ^{2}\nabla^{2}}{2m} \nonumber \\
&+&V_{opt}(\mathbf{r})+\frac{g_{e}}{2}|\psi(
\mathbf{r},\tau )|^2\Bigg]\psi (\mathbf{r},\tau ) \label{Action}
\end{eqnarray}
is the action functional of
$\left[\psi^{*}\left(\mathbf{r}
,\tau\right),\psi\left(\mathbf{r},\tau\right)\right]$ which collectively
denote the complex functions of space and imaginary time $\tau$. Here, $g_{e}$ abstractly stands for the
two-body coupling constant in an axial optical confinement. Using the path-integral approach \cite{Popov}, one finds within the tight-binding approximation and Bogoliubov theoretical framework \cite{Zhou} the ground state energy $E_g$,
\begin{eqnarray}
\frac{E_g}{V}=\frac{1}{2}\tilde{g}_en^2\Bigg[&1&+\frac{m\tilde{g}_e}{2\pi^2\hbar^2d}F\left(\frac{2t}{\tilde{g}_en}\right)
\Bigg],\label{Eg}
\end{eqnarray}
with
\begin{eqnarray}\label{F}
F(x)&=&\frac{(x+1)}{2}\left[\left(3x+1\right)\arctan\left(\frac{1}{\sqrt{x}}\right)-3\sqrt{x}\right]\nonumber\\
&-&\frac{\pi}{2}\ln \left[\frac{x}{2x+1+2\sqrt{x\left(x+1\right)}}\right]-\pi\text{arcsinh}\left(\sqrt{x}\right)\nonumber\\
&+&2\int_{0}^{\sqrt{x}}\frac{\tan^{-1}(z)}{z}dz.
\end{eqnarray}
Here, $t$ denotes the tunneling rate, $n$ refers to the condensate density, and $\tilde{g}_e=g_e\left[d\int_{-d/2}^{d/2}w^4(z)dz\right]=g_e({d}/{\sqrt{2\pi}\sigma})$, where $w\left(u\right)=\exp\left[-u^2/2\sigma^2\right]/\pi^{1/4}\sigma^{1/2}$ is a variational Gaussian anstaz and the ratio $d/\sigma\simeq \pi s^{1/4}\exp\left(-1/4\sqrt{s}\right)$ minimizes the free energy functional with respect to $\sigma$ \cite{Orso}.

The ground state energy in Eq. (\ref{Eg}) experiences a lattice-induced dimensional crossover governed by the parameter $2t/\tilde{g}_en$. In the limit $2t/\tilde{g}_en\gg 1$, one finds the system exhibiting anisotropic 3D behavior and $F(x)\simeq 32/15 \sqrt{x}$. Whereas, the extreme $2t/\tilde{g}_{e}n\ll 1$ corresponds to the quasi-2D regime where axial atomic motion is frozen to zero-point oscillations and $F(x)=\frac{\pi}{4}-\frac{\pi}{2}\log x$ is approached exactly. We stress that above analysis is justified by the weak but nonegligible tunneling effect guaranteed by $1/N_t\leq t/\tilde{g}_en$ with $N_t$ being the number of atoms per optical well.

Upon transiting from 3D to quasi-2D regime $1/N_t\leq 2t/\tilde{g}_{e}n\ll 1$, the axial optical lattice imprints its effect in binary atomic collisions via strongly restricting axial kinematics that manifests as a 2D character of the relative motion of colliding atoms at large separation. This gives rise to an effective coupling constant sensitive to the lattice parameter which is approximately given by \cite{Wouters, Petrov}
\begin{equation}\label{gh}
g_{e}=\frac{2\sqrt{2\pi}\hbar^2d}{m}\frac{1}{a_{2D}/a_{3D}+(1/\sqrt{2\pi})\ln{\left[1/n_{2D}a_{2D}^2\right]}},
\end{equation}
where $n_{2D}=nd$ is the surface density and $a_{2D}=\sigma$ is the effective 2D
scattering length. The logarithmic density-dependent term in Eq. ({\ref{gh}}) is typical of 2D many-body effects, whose relative importance is governed by the ratio $a_{2D}/a_{3D}$ which henceforth controls the dimensional crossover in hard-core interactions \cite{Petrov}. For $a_{2D}/a_{3D}\gg 1$, one finds pure 3D collision and Eq. ({\ref{gh}}) converges to the lattice-renormalized density independent coupling constant $\tilde{g}=\frac{d}{\sqrt{2\pi}\sigma}\frac{4\pi\hbar^2a_{3D}}{m}$. Whereas in the opposite limit $a_{2D}/a_{3D}\ll 1$, 2D scattering dominates which leads to $g_e=g_{2D}d$ with
$g_{2D}=(4\pi \hbar ^{2}/m)/\ln\left(1/n_{2D}a_{2D}^{2}\right)$ being the density-dependent effective coupling constant peculiar to pure 2D Bose gases. Hereafter, we focus on what we shall call the weak quasi-2D regime where the small parameter $a_{2D}/a_{3D}$ measures the deviation from pure 3D scattering due to emerging 2D features. Linearly expanding Eq. (\ref{gh}) with respect to ${a_{3D}}/{a_{2D}}$, one obtains
\begin{equation}\label{Ge1}
\tilde{g}_e=\tilde{g}\left[1-\frac{1}{\sqrt{2\pi}}\frac{a_{3D}}{a_{2D}}\ln\left(\frac{1}{n_{2D}a_{2D}^2}\right)\right],
\end{equation}
where the term showing the logarithmic dependence on the gas parameter constitutes the leading 2D correction to 3D coupling constant $\tilde{g}$.

The $\mu_{Q2D}$ can now be readily determined via $\mu=\partial E_g/\partial N$ from Eq. ({\ref{Eg}}), together with proper asymptotic analysis. We first note that in the asymptotic 3D regime where $2t/\tilde{g}n\gg 1$ and $a_{3D}/a_{2D}\gg 1$, our analytical solution
$\mu=\tilde{g}n\left[1+({32m^{*}}/{3\sqrt{\pi}m})\sqrt{a_{3D}^3n} \right]$ is consistent with the 3D Lee-Huang-Yang (LHY) result \cite{Yang}; whereas, in the opposite pure 2D limit where $2t/\tilde{g}n\ll 1$ and $a_{3D}/a_{2D}\ll 1$, our asymptotical result for the chemical potential of a 2D Bose gas
$\mu=\frac{4\pi\hbar^2n_{2D}/m}{|\ln n_{2D}a^2_{2D}|}
\Big[1-\frac{\ln\left(\ln\left(1/n_{2D}a^2_{2D}\right)\right)-B}{\ln\left(1/n_{2D}a^2_{2D}\right)}
-\frac{\ln\left(\ln\left(1/n_{2D}a^2_{2D}\right)\right)-B}{\ln^2\left(1/n_{2D}a^2_{2D}\right)}\Big]$
with $B=1-\ln \left(mt/n_{2D}2\pi \hbar ^{2}\right)$ stands in good agreement with Ref. \cite{Mora}. Thence, after applying similar schemes to the weak quasi-2D regime where $2t/\tilde{g}n\ll 1$ and $a_{3D}/a_{2D}$ is small, one finds
\begin{equation}\label{K}
\mu_{Q2D}\!=\!\tilde{g}n\left[1\!+\!\frac{1}{\sqrt{2\pi}}\frac{a_{3D}}{a_{2D}}\left(\frac{1}{2}\!-\!\ln\left(\frac{1}{n_{2D}a^2_{2D}}\right)\right)\right].
\end{equation}
Rewriting $\mu_{Q2D}=\tilde{g}n\left[1+k_{2D}\left(n\right)\right]$, we thus identify $k_{2D}(n)=a_{3D}/(\sqrt{2\pi}{a_{2D}})\left[1/2-\ln\left({1}/{n_{2D}a^2_{2D}}\right)\right]$ as the first correction to the 3D mean-field (MF) equation of state arising from the 2D effect.

From Eq. ({\ref{K}}), the equation for the 3D ground state density can be solved by iteration yielding
\begin{equation}\label{GS}
{n}({\bf r})=n_{TF}-\frac{1}{\sqrt{2\pi}}\frac{a_{3D}}{a_{2D}}\left[\frac{1}{2}+\ln\left(\frac{1}{dn_{TF}a_{2D}^2}\right)\right]n_{TF},
\end{equation}
with $n_{TF}\left({\bf r}\right)=\left(\mu_0-V_{ext}\left({\bf r}\right)\right)/\tilde{g}$ being the 3D TF density. Eq. ({\ref{GS}}) clearly shows that, because of the weak coupling between adjacent wells, the 2D corrections in the local chemical potential in Eq. ({\ref{K}}) are transferred to the 3D stationary shape of cloud.

Substituting Eqs. (\ref{K}) and (\ref{GS}) into Eq. (\ref{HD33}) and only retaining terms linear in $k_{2D}(n)$, we obtain
\begin{equation}\label{HDF}
m\omega^2\delta n\!+\!\tilde{\nabla}\cdot\left(\tilde{g}n_{TF}\tilde{\nabla}\delta n\right)\!=\!-\!\tilde{\nabla}^2\left(\tilde{g}n^2_{TF}\frac{\partial k_{2D}}{\partial n_{TF}}\delta n\right).
\end{equation}
Equation ({\ref{HDF}}) in the absence of $k_{2D}$ is just the familiar 3D hydrodynamic equation in the presence of 1D optical lattice \cite{Kramer}.  Against this background, the addition of terms on the right side of Eq. (\ref{HDF}) presents a perturbation. The resulting fractional frequency shift, to the leading order, is given by
\begin{equation}\label{Shift}
\frac{\delta \omega}{\omega}=-\frac{\tilde{g}}{2m\omega^2}\frac{\int d^3{\bf r}\tilde{\nabla}^2\delta n^{*}\left(n^2_{TF}\frac{\partial k_{2D}}{\partial n_{TF}}\delta n\right)}{\int d^3{\bf r}\delta n^{*}\delta n},
\end{equation}
where integrals extend to the region where $n_{TF}$ is positive \cite{Pitbeyond}. An important feature in Eq. (\ref{Shift}) is the dependence of $\delta\omega/\omega$ on the derivative $\partial k_{2D}/\partial n$ rather than $k_{2D}(n)$. The consequence is that the leading order correction arising from 2D effect to 3D MF collective frequency shows no logarithmic density dependence.

According to Eq. ({\ref{Shift}}), the surface modes that satisfy $\tilde{\nabla}^2\delta n= 0$ are not perturbed by the 2D effect in the vicinity of 3D regime. Hence, in order to observe dimensional effects, one has to focus on small compressional oscillations. Our primary mode of interest is the transverse breathing mode in a very elongated trap ($\sqrt{m/m^{*}}\omega_z/\omega_\perp\ll 1$).  Substitutions of $\delta n({\bf r})\sim r_\perp^2-R_{TF}^2/2$ with $R_{TF}=\sqrt{2\mu_0/m\omega_\perp^2}$ being the transverse TF radius and $\omega=2\omega_{\perp}$ into Eq. ({\ref{Shift}}) yield the fractional shift
\begin{equation}\label{BreathW}
\frac{\delta \omega}{\omega }=\frac{1}{4\sqrt{2\pi}}\frac{a_{3D}}{a_{2D}}.
\end{equation}
Equations ({\ref{Shift}}) and ({\ref{BreathW}}) consist of the major results of this paper. In typical experiments to date \cite{Du}, the relevant parameters are given by 3D scattering length $a_{3D}=5.31nm$ and the lattice period $d=297.3nm$. The frequency shift in Eq. (\ref{BreathW}) can be reached $\sim 0.48\%$ for $s=4$. Given an accuracy of $\sim 0.3-0.4\%$ in measuring collective frequencies within current facilities \cite{CEM}, the 2D correction to the transverse breathing mode is well in reach in relevant experiment conditions. Moreover, this effect can be enhanced via adjusting lattice parameter and using Feshbach resonance. We have also taken a look at the lowest compression mode in a disk-like geometry ($\sqrt{m/m^{*}}\omega_z/\omega_\perp\gg 1$), which is along the axial direction with the zeroth order dispersion given by $\omega=(\sqrt{{m}/{m}^{*}})\sqrt{3}\omega_z$ and density oscillations of the form $\delta n({\bf r})\sim z^2-Z_{TF}^2/3$, where $Z_{TF}=\sqrt{2m^{*}\mu_0/m\omega_z^2}$ is the TF radius along the axial direction. Straightforward calculations yield $\frac{\delta\omega}{\omega} = \frac{1}{6\sqrt{2\pi}}\frac{m^{*}}{m}\frac{a_{3D}}{a_{2D}}$, showing an amplified 2D effect due to the increased inertia along the direction of the laser.

The frequency shift in Eq. (\ref{BreathW}) should be compared with other relevant corrections in actual experiments, like finite size, nonlinearity, thermal effects and vortex. Finite size effects originate from kinetic energy pressure typically ignored in the TF scheme. Its consequence on the transverse breathing mode can be analyzed from Eq. (\ref{Breath}), averaged over the many-body wavefunctions in accordance with the experimental measurements. For very elongated geometry where nearly spatial invariance in axial direction implies $E_{kinz}=0$, calculations by GP theory directly give $\omega=2\omega_{\perp}$, showing no finite size effect. In general case where $\omega_z$ is finite and $E_{kinz}\neq 0$, a sum rule approach within single-mode approximation gives $\omega=\sqrt{m_3/m_1}$ with $m_3=\left(8\hbar^4/m^2\right)\left(E_{kin\perp}+E_{ho\perp}+E_{int}\right)$ and $m_1=(2\hbar^2/m)N\langle F_B\rangle$ respectively being the cubic energy weighted and the energy weighted moments of the dynamic structure factor $m_p=\sum_{n}|\langle 0|F_B|n\rangle|^2(\hbar \omega_{n0})^{p}$. Using the viral identity $E_{kin\perp}-E_{ho\perp}+E_{int}=0$ for the ground state in GP description, one again finds $\omega=2\omega_{\perp}$ unaffected by the finite size effect.

In actual experiments where the amplitude of oscillation can not be arbitrarily small, effects of nonlinearity can shift the collective frequency approximately by $\delta\omega/\omega=A^2\delta$ \cite{Dalfovo}. Here $A$ is the fractional oscillating amplitude of the harmonically confined atomic cloud which can be tuned to less than $10\%$. In our case, the coefficient $\delta$ is calculated as $\delta=\frac{5}{2}\lambda^2\frac{(q_{-}-2)(q_{+}-4)(q_{-}-5)}{(4q_{+}-q_{-})(q_{-}-q_{+})^2}\left[-1+\frac{15}{4}\frac{\lambda^2}{q_{+}^2}\right]
-\frac{15}{16}\frac{1}{(q_{-}-q_{+})^2}\left[-q_{+}+2\lambda^2q_{+}-9\lambda^2+8\right]^2-\frac{9}{4}\frac{(q_{-}-4)}{q_{+}(q_{+}-q_{-})}
-\frac{3}{20}\frac{q_{+}-3}{q_{+}-q_{-}}\left[-10\lambda^2q_{+}+37\lambda^2+11q_{+}-54\right]$ with $q_{\pm}=2+\frac{3}{2}\lambda^2\mp \frac{1}{2}
\sqrt{9\lambda^4-16\lambda^2+16}$, which asymptotically varnishes in the limit $\lambda=\omega_z/\omega_\perp\rightarrow 0$. The effect of nonlinearity is therefore very small.

The consequence of thermal effects on collective excitations has been experimentally investigated in Ref. \cite{Chevy}. The observed unusually small damping rates and frequency shifts in the transverse breathing mode of an elongated condensate have been attributed to an accidental degeneracy between the condensate and thermal cloud oscillation frequencies \cite{Jackson}. This accidental suppression of Landau damping may offer an experimental control of the thermal effect on the frequency shift.

The issue of vortex is closely related to operational anisotropy in excitation schemes. In actual experiments, the breathing mode is generally excited via quenching the transverse harmonic trap frequencies {\it{in phase}} by $\delta\omega_{x(y)}=\delta\omega\ll \omega_{\perp}$. However, this scheme will be perturbed by an out-of-phase operation with $\delta\omega_x=\delta \omega$ and $\delta\omega_y=\delta\omega^{'}$, and the resulting excitation operator $F= m\omega_{\perp}(\delta\omega+\delta\omega^{'})/2\times F_{B}+m\omega_{\perp}(\delta\omega-\delta\omega^{'})/2\sum_{i}r_i^2\left(Y_{2,2}+Y_{2,-2}\right)$ ($Y_{lm}$ being spherical harmonics) gives rise to additional excitations of quadrupole modes. Such situation is further spoiled by the  presence of quantum vortex which splits the $m=\pm 2$ quadrupole modes by approximately $\omega_{+}-\omega_{-}=\frac{7\omega_{\perp}\kappa}{\lambda^{2/5}}\left(15\frac{N\tilde{a}_{3D}}{a_{\perp}}\right)$ \cite{Zambelli}. We note that this split becomes infinite when $\lambda\rightarrow 0$, suggesting an important point of observing our result in Eq. ({\ref{BreathW}}) is to avoid the anisotropy in perturbation schemes and presence of a vortex.

In conclusion, our results show that 2D many-body effects can be visible in the frequency shift of the transverse breathing mode for an optically trapped Bose gas transiting from 3D to quasi-2D regime. Observing dimensional effect directly would present an important achievement in revealing the interplay between dimensionality and quantum fluctuations in low-dimensional strongly correlated quantum systems associated with BEC.

Noting that compared to the fruitful work on a BEC along the 3D, quasi-1D and 1D dimensional crossover \cite{Quasi1D}, much less have been reported in the transition region from the 3D to quasi-2D and 2D regimes where many unique phenomena are known to arise. More comprehensive study along this line is highly desirable.

We thank Biao Wu for helpful discussions. This work is supported by the NSF of China (Grant No. 11004200) and IMR SYNL-TS K\^e Research Grant.

\end{document}